\documentstyle[11pt,newpasp,twoside]{article}
\def\edcomment#1{\iffalse\marginpar{\raggedright\sl#1\/}\else\relax\fi}
\reversemarginpar

\begin{document}
\title{Possible Transiting Planet Candidates from the EXPLORE Project}

\author{G.\ Mall\'en-Ornelas}
\affil{Princeton University Observatory, Peyton Hall, Princeton, NJ 08544 and
P.\ Universidad Cat\'olica de Chile, Casilla 306, Santiago 22, Chile}
\author{S.\ Seager}
\affil{Institute for Advanced Study, Einstein
Drive, Princeton, NJ 08540 and The Carnegie Institution of Washington,
Dept. of Terrestrial Magnetism, 5241 Broad Branch Rd. NW, Washington,
DC 20015 }
\author{H.\ K.\ C.\ Yee}
\affil{Department of Astronomy and Astrophysics, University of Toronto,
60 St.\ George St., Toronto, ON M5S 3H8, Canada}
\author{Michael D.\ Gladders}
\affil{Department of Astronomy and Astrophysics, University of Toronto,
60 St.\ George St., Toronto, ON M5S 3H8, Canada and The Carnegie
Observatories, 813 Santa Barbara St., Pasadena, CA 91107}
\author{T.\ M.\ Brown}
\affil{High Altitude Observatory/National Center for Atmospheric
Research, P.O.\ Box 3000, Boulder, CO 80307}
\author{D.\ Minniti}
\affil{Departamento de Astronom\'{\i}a y Astrof\'{\i}sica, Pontificia
Universidad Cat\'olica de Chile, Casilla 306, Santiago 22, Chile}
\author{S.\ L.\ Ellison}
\affil{European Southern Observatory, Casilla 19001, Santiago, Chile}
\author{G.\ M.\ Mall\'en-Fullerton}
\affil{Universidad Iberoamericana, Prolongaci\'on Paseo de la Reforma
880, 01200 M\'exico, D.F., M\'exico}

\begin{abstract}

The EXPLORE Project is a series of searches for transiting extrasolar
planets using large-format mosaic CCD cameras on 4-m class telescopes.
Radial velocity follow-up is done on transiting planet candidates with
8--10m class telescopes.  We present a summary of transit candidates
from the EXPLORE Project for which we have radial velocity data.
\end{abstract}

\section{Introduction}

The EXPLORE (EXtrasolar PLanet Occultation REsearch) Project is a
series of searches for transiting extrasolar planets orbiting Galactic
plane stars using 4-m class telescopes.  As an integral part of the
search strategy, radial velocity (RV) follow-up observations for mass
confirmation are done on 8--10m class telescopes.  In June 2001, we
used the CTIO 4-m telescope for 11 nights (6 clear) to observe the
EXPLORE~I field, located at $l=-28, b=-3$ (Mall\'en-Ornelas et al.\
2002).
The best 37,000 light curves were examined, and RV follow-up of three
planet candidates was done on the VLT in September 2001
(Mall\'en-Ornelas et al., in prep.).  In December 2001 we used the
3.6-m CFHT for 16 nights (14 clear) to observe the EXPLORE~II field,
located at $l=203, b=0.7$ (Yee et al., in prep.).  The best 12,000
light curves were examined, and RV follow-up of 2 planet candidates
and 2 additional eclipsing systems was done at Keck in February 2002
(Mall\'en-Ornelas et al., in prep.).  Here we present a summary of the
transiting planet candidates for which we have RV follow-up.

\section{What Makes a Good Transiting Planet Candidate?}

Light curves with very precise photometry and good time sampling can
be used to determine the eclipse shape and select the best transiting
planet candidates with minimal contamination from other systems that
can mimic the transit signature.  In general terms, the best planet
candidates will have very shallow eclipses ($\la$2--3\%, implying a
small companion) of short duration ($\la$3h for a 3--4 day period
transit, implying a dwarf star), with a clear flat bottom, and
preferrably with a steep ingress/egress (i.e., a ``box shape'').  
For a full discussion on candidate selection and possible contaminants
see Seager \& Mall\'en-Ornelas (2002) and Mall\'en-Ornelas et al.\
(2002).  A mass measurement from RV observations is required in order
to confirm the presence of a planet, as opposed to a brown dwarf or a
stellar companion.  In particular, the RV data should show only one
velocity peak (indicative of the presence of a single star), and
ideally result in a mass detection (with $M_{planet} < 13 M_J$).  Note
that in the absence of an actual mass detection, a good upper limit of
1--2 $M_J$ may be used to make the case for a planet {\it as long as
the eclipses can be confirmed with certainty, and possible
contaminants can be ruled out with confidence}.

\section{Possible Planet Candidates from the EXPLORE Project}

A total of three planet candidates from the EXPLORE~I search were
followed-up with the VLT+UVES in September 2001.  For the EXPLORE~II
search, two planet candidates as well as two (brigher) stars with
deeper eclipses were observed with Keck+HIRES in February 2002.  Based
on the window function of our observations, on the number of light curves
examined, assuming that 0.75\% of single stars have a close-in giant
planet, and adopting a binary fraction of 1/2, we expect to find 1
planet in the EXPLORE~I search and 1--2 planets in the EXPLORE~II
search.  Here we summarize our preliminary findings on the five planet
candidates for which we have conducted RV follow-up observations.  The
essential information is presented in tables 1 and 2, and a brief
discussion is given for each candidate.  A full description of our
results will be presented in two future papers (Mall\'en-Ornelas et
al., in prep.).  Light curves and preliminary RV plots for some of the
systems discussed here can be found in Yee et al.\ (2002).

\subsection{Summary of Possible Planet Candidates}

Table~1 lists the I and V magnitudes, eclipse depth, and period for
the five transiting planet candidates with RV follow-up.  Note that in
some cases it is possible that the period listed is a multiple of the
true period, since additional intervening eclipses may have occurred
during the day.  Table~2 gives the eclipse and RV characteristics of
the transit candidates, which can be used to determine whether each
system is a good planet candidate.  {\it As mentioned in \S~2, the
best planet candidates have ``box-shaped'' short-duration eclipses
with clear flat bottoms.  In the case of a planet the RV should show
only one velocity peak in the cross-correlation, and any RV variations
should be of very small amplitude.}  For some of the EXPLORE data, poor S/N in
the light curve prevented us from ascertaining the eclipse shape in
some cases, marked with a ``?'' in Table~2.  Bad weather during the RV
follow-up resulted in poor phase coverage for EX2c11s4809, which
unfortunately prevented us from measuring a mass for this otherwise
promising planet candidate.

\begin{center}
\begin{tabular}{|l|cc|cc|c|}\hline \hline
\multicolumn{6}{|c|}{Table 1: General characteristics for transit candidates with RV follow-up}\\
\hline
Star ID&I&V &\# of Eclipses&Eclipse Depth&Period\\
&(mag)&(mag)&&(mag)&(days)\\
\hline
EXP1c02s46830&16.1&17.7&1\footnote{there is a possible second eclipse with low S/N}&0.015&?\\
EXP1c01s52805&16.2&17.9&2&0.03&2.2\\
EXP1c07s18161&17.6&19.4&2&0.025&3.8\\
\hline
EXP2c11s4809&18.3&20.0&3&0.017&3.0\\
EXP2c10s5069&18.4&20.1&4&0.008&4.0\\
\hline\hline
\multicolumn{6}{l}{$^1$ there is a possible second low-S/N eclipse in this object}
\end{tabular}
\end{center}

\begin{center}
\begin{tabular}{|l|ccc|cc|c|}\hline \hline
\multicolumn{7}{|c|}{Table 2: Eclipse and RV characteristics for transit candidates}\\
\multicolumn{7}{|c|}{with RV follow-up}\\
\hline
Star ID&\multicolumn{3}{c|}{Eclipse Characteristics}&\multicolumn{2}{c|}{RV Characteristics }&Possible\\
&Flat?&Boxy?&Short?&Peaks&Shifts&Planet?\\
\hline
EXP1c02s46830&?&N&Y&1&Y&?\\
EXP1c01s52805&Y&N&Y&2&Y&No\\
EXP1c07s18161&?&?&Y&1&N&Yes\\
\hline
EXP2c11s4809&Y&Y&Y&1&?&Yes\\
EXP2c10s5069&?&?&Y&1&N&Yes\\
\hline\hline
\end{tabular}
\end{center}

\subsection{Individual systems}

\noindent $\bullet$ 
{\bf EXP1c02s46830.}  Only one clear eclipse was detected, and the long
ingress and egress duration make it an unlikely planet candidate.  A
steady decrease in RV of $\sim$500 m/s was detected over the 8 days
spanned by the observations.  However, a planet cannot be ruled out
until a period is measured (either from the eclipses or the RV
variations) which will enable a mass determination.

\noindent $\bullet$ 
{\bf EX1c01s52805.}  The RV data for this object had two velocity
peaks, revealing a system composed of a close eclipsing binary, plus a
brighter superimposed star with no significant RV shifts.  Note that
the eclipses for this system have clear flat bottoms but {\it are not}
``box-shaped''.  Although it is possible for a planet crossing close
to the stellar limb to produce transits with a long ingress and
egress, such planet transits are much less common than ``box-shaped''
transits (Seager \& Mall\'en-Ornelas 2002).

\noindent $\bullet$ {\bf EXP1c07s18161.}  This system has 2 detected
eclipses, though the S/N is not high enough to ascertain eclipse shape
with confidence.  There is only one RV peak, and we place a
preliminary upper mass limit of 2--3 $M_J$.  Thus, this is a very
promising planet candidate.  To definitely establish this system as a
planet, we require further photometric observations to confirm the
eclipses, and ideally more RV data 
to obtain an actual mass detection.

\noindent $\bullet$ 
{\bf EXP2c11s4809.} This system has very clear flat-bottomed
eclipses, and only one RV peak.  However, due to poor weather we were
unable to obtain enough phase coverage to place a useful mass limit.
Further RV observations leading to a mass measurement are required to
establish the nature of this system.

\noindent $\bullet$ 
{\bf EXP2c10s5069.} We detected 4 eclipses, but all with very low
S/N so the eclipse shape could not be determined with certainty.  Only
one RV peak was found, with no detected shifts.  To establish this
system as a planet we need more RV data to attempt a mass detection,
as well as photometric observations to increase the S/N of the
eclipses.

\section{Summary and Conclusions}

We have presented a status report on the transit candidates from the
EXPLORE Project's first two surveys.  Out of three systems with RV
follow-up in the EXPLORE~I search, one was found to be a close stellar
binary plus a blended star, a second one is an unlikely planet
candidate that nonetheless has not been entirely ruled out as a
planet, and a third one is a promising planet candidate based on an
upper mass limit.  The two candidates with RV follow-up in the
EXPLORE~II search remain good planet candidates, although more RV data
are required for both systems, and more photometric data are also
desirable for EXP2c10s5069, which has extremely shallow eclipses
detected with low S/N.

We have shown that it is possible to produce a clean set of planet
candidates from a deep transit search and place meaningful mass limits
(2--3 $M_J$) on the companions via follow-up RV observations.  We
emphasize that in order to find one good planet candidate, it is
necessary to monitor several thousands of stars with good time
coverage, high photometric precision, and time sampling which is good
enough to determine the eclipse shape.

\end{document}